\documentclass[
tightenlines,
nofootinbib,
twocolumn,
 amsmath,amssymb,
 aps,
]{revtex4-1}
\usepackage{amsmath}
\usepackage{amssymb}
\usepackage{bm}
\usepackage{color,graphicx}
\usepackage{slashed}
\usepackage{comment}

\newcounter{comment}

\begin{document}

\title{Novel Rosenbluth Extraction Framework for Compton Form Factors from Deeply Virtual Exclusive Experiments}

\author{Brandon Kriesten} 
\email{btk8bh@virginia.edu}
\affiliation{Department of Physics, University of Virginia, Charlottesville, VA 22904, USA.}

\author{Simonetta Liuti} 
\email{sl4y@virginia.edu}
\affiliation{Department of Physics, University of Virginia, Charlottesville, VA 22904, USA.}

\author{Andrew Meyer}
\email{ajm5an@virginia.edu}
\affiliation{Department of Physics, University of Virginia, Charlottesville, VA 22904, 
USA.}%


\begin{abstract}
We use a generalization of the Rosenbluth separation method for a model independent simultaneous extraction of the  Compton Form Factors ${\cal H}$ and  ${\cal E}$, 
from virtual Compton scattering data on an unpolarized target. A precise evaluation of ${\cal H}$ and ${\cal E}$, enabled by the proposed method, is the first step towards pinning down the the distribution of angular momentum inside the proton.
\end{abstract}

\maketitle

 Deeply virtual Compton scattering (DVCS) has been identified as the cleanest probe to study the 3D structure of the proton, as well as its mechanical properties including angular momentum \cite{Ji:1996ek,Ji:1996nm} and the pressure and shear forces \cite{Polyakov:2002yz} carried by quarks and gluons (see reviews in \cite{Diehl:2003ny,Belitsky:2005qn,Kumericki:2016ehc}). Information on these properties is encoded 
in the Generalized Parton Distributions (GPDs) which  parametrize simultaneously the quark-proton correlation function and, integrated over the longitudinal variable, the QCD energy momentum tensor. An important distinction with deeply virtual inclusive processes is that GPDs enter the cross section at the amplitude level, convoluted with the hard scattering matrix elements, embedded into both the real and imaginary  Compton Form Factors (CFFs). One has, therefore, a total of eight CFFs constructed from the GPDs $H, E, \widetilde{H}, \widetilde{E}$. 

Current experimental programs at Jefferson Lab and at the newly planned Electron Ion Collider (EIC) are expected  to pose stringent constraints on the CFFs  
in a wide kinematic range in $Q^2$, the four-momentum transfer squared between the initial and final electrons, $t$, the four-momentum transfer squared between the initial and final protons, and  Bjorken $x_{Bj}$ (the latter is related to the skewness parameter, $\xi$ \cite{Ji:1996ek,Radyushkin:1997ki}.
Setting $Q^2 >> t$  in the multi-GeV region, in particular, provides a scale where QCD factorization is predicted to hold    \cite{Collins:1996fb,Collins:1998be,Ji:1997nk,Ji:1996nm,Ji:1998xh,Blumlein:1997pi,Blumlein:1999sc}.
%
%
A clear-cut, model independent extraction of the CFFs from from experiment poses, however, an unprecedented challenge for data analyses (see {\it e.g.} Ref.\cite{Moutarde:2009fg,Guidal:2013rya}). 
A major hindrance has been that 
the coefficients multiplying the various CFFs combinations cannot be straightforwardly organized into a form reflecting the QCD twist expansion, as kinematic power corrections occur \cite{Braun:2014sta}.  

In this Letter we argue that  
most of the difficulties in the CFF extraction stem from 
having overlooked important dynamical aspects of deeply virtual exclusive reactions specific to the description of coincidence scattering processes originally displayed in \cite{Arens:1996xw,Diehl:2005pc,Vanderhaeghen:1998uc}.
Left without identifying its primary electric, magnetic and axial current components, the cross section has been cast 
in a harmonic expansion in the azimuthal angle between the lepton and hadron planes \cite{Belitsky:2001ns}. The harmonic expansion introduces cumbersome kinematic coefficients (see {\it e.g.} Ref.\cite{Belitsky:2010jw} and the Appendix of Ref.\cite{Moutarde:2009fg}). To simplify these expressions, various approximations have been introduced that affect a clear-cut extraction of CFFs from data. Most importantly, this formalism lacks a criterion to understand and interpret the GPD content of any particular beam-target polarization configurations. 

Conversely, by casting the cross section into a form that fits our physical knowledge from coincidence reactions, in terms of  generalized magnetic, electric and axial form factors allows us to extract for the first time the CFFs ${\cal H}$ and ${\cal E}$ simultaneously, from a linear fit, using unpolarized data only. A corollary to this extraction is that the contribution of $\widetilde{\cal H}$ is suppressed, as expected, since this term would be parity violating in the limit of elastic scattering, or for the emission of a photon with zero momentum.  


In Refs.\cite{Kriesten:2019jep,Kriesten:2020wcx} we 
introduced a new framework where the Bethe Heitler (BH) and the  deeply virtual Compton scattering (DVCS), interference contributions are cast in terms of generalized electric, magnetic and axial form factors as, 
%
\begin{figure*}
    \centering
    \includegraphics[width=7cm]{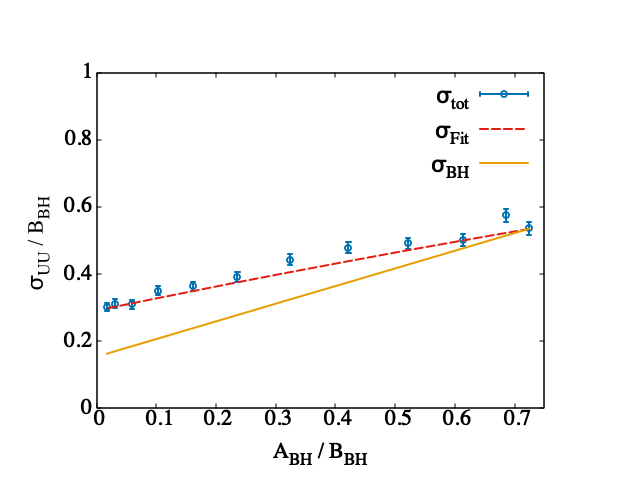}
    \hspace{-0.5cm}
    \includegraphics[width=7cm]{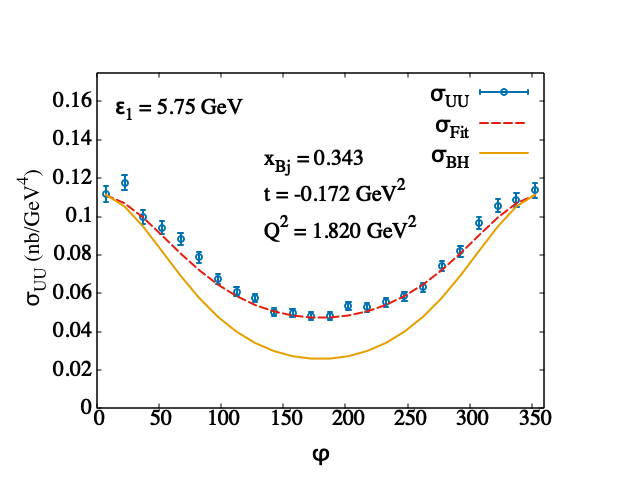}
    \caption{{\it Left}: reduced cross section, $\sigma_{UU} t^2/\Gamma B_{BH}$ plotted vs $A_{BH}/B_{BH}$ in the experimental kinematic bin: $x_{Bj}=0.343$, $t=-0.172$ GeV$^2$, $Q^2 = 1.82$ GeV$^2$, \cite{Defurne:2015kxq}; {\it Right}: the cross section, $\sigma_{UU}$, plotted vs $\phi$. The orange lines on both sides correspond to the BH calculation using the values of the form factors from Ref.\cite{Kelly:2004hm}. The data deviate of from this straight line due to the effect of DVCS. We also show the complete result for the cross section obtained with the CFFs extracted in this paper (red dashed curves).}
    \label{fig:BH}
\end{figure*}
%

\vspace{0.2cm}
\begin{widetext}
\begin{eqnarray}
\label{eq:siguuBH}
 \frac{d^4\sigma_{UU}^{BH}}{d Q^2 d x_{Bj} d t d \phi} &=& 
 \label{eq:BH}
\frac{\Gamma}{t^2}  \Big\{A_{BH} \left[F_1^2(t) + \tau F_2^2(t) \right]+ B_{BH} \tau G_M^2(t)  \Big\} 
\\
\label{eq:siguuI}
\frac{d^4\sigma_{UU}^{\cal I}}{d Q^2 d x_{Bj} d t d \phi}  &= & 
 \frac{e_l \Gamma}{ Q^2 \mid t \mid} \Big\{ A^{\cal I}   \Re e \big[F_1(t) \mathcal{H}(\xi,t) + \tau F_2(t)  \mathcal{E}(\xi,t) \big]   + B^{\cal I}    G_M(t)  \Re e \big[ \mathcal{H}(\xi,t)+ \mathcal{E}(\xi,t) \big]
  +  C^{\cal I}   
G_M(t)  \Re e \mathcal{\widetilde{H}}(\xi,t) \Big\}
 \\
  \label{eq:sigLUI}
\frac{d^4\sigma_{LU}^{\cal I}}{d Q^2 d x_{Bj} d t d \phi} &= & \, \frac{e_l \Gamma}{ Q^2 \mid t \mid} 
\Big\{ A^{\cal I}   \Im m \big[F_1(t) \mathcal{H}(\xi,t) + \tau F_2(t)  \mathcal{E}(\xi,t) \big]   + B^{\cal I}    G_M(t)  \Im m \big[ \mathcal{H}(\xi,t)+ \mathcal{E}(\xi,t) \big]
  +  C^{\cal I}   
G_M(t)  \Im m \mathcal{\widetilde{H}}(\xi,t)  \Big\} \nonumber \\
\end{eqnarray}
\end{widetext}
where $t=(p'-p)^2$ is the four-momentum transfer squared between the initial ($p$) and final ($p'$) proton; $\xi\approx x_{Bj}/(2-x_{Bj})$; $\tau = -t/(4M^2)$, $M$ being the proton mass; $\phi$ is the azimuthal angle between the lepton and hadron scattering planes; $e_l$ is the lepton charge; $F_1$ and $F_2$ are the Dirac and Pauli form factors, ($G_M = F_1+F_2$, and $G_E = F_1 - \tau F_2$); the kinematic factor $\Gamma$ accounts for the flux factor; $A_{BH}$ and $B_{BH}$, $A^{\cal I}$, $B^{\cal I}$, and  $C^{\cal I}$ are kinematic coefficients whose detailed expressions 
are given in Ref.\cite{Kriesten:2019jep,Kriesten:2020wcx}. 
\footnote{In DVCS we the CFFs for a generic GPD $F(x,\xi,t)$ are defined as: \[{\cal F} = \int^1_{-1} dx \Big[ \frac{1}{x -\xi- i \epsilon} - \frac{1}{x + \xi - i \epsilon} \Big] F(x,\xi,t) ,\] where $\xi \approx x_{Bj}/(2 - x_{Bj})$. }

The apparently simple notion of re-parametrizing the cross section as in the equations above revolutionizes the way we look at deeply virtual exclusive experiments analyses, uniquely leading to a major improvement in the precision extraction of CFFs from data. High precision is what is needed to turn the extraction of 3D images of the proton into a reality. 
\begin{figure*}
    \centering
    \includegraphics[width=7cm]{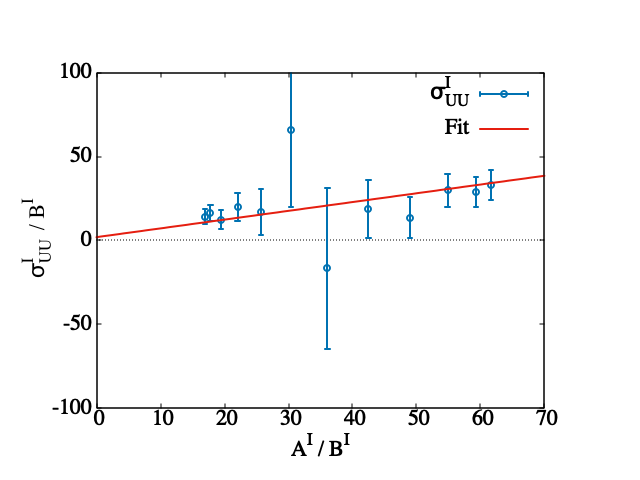}
    \hspace{-1cm}
    \includegraphics[width=7cm]{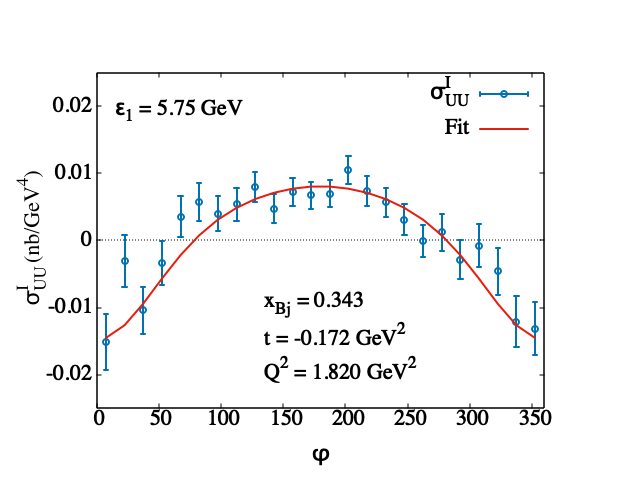}
    \caption{{\it Left}: reduced BH-DVCS interference contribution to the unpolarized cross section, $\sigma_{UU}^{\cal I} Q^2 |t |/B^{\cal I}$ plotted vs $A^{\cal I}/B^{\cal I}$ for the  experimental kinematic bin: $x_{Bj}=0.343$, $t=-0.172$ GeV$^2$, $Q^2 = 1.82$ GeV$^2$, \cite{Defurne:2015kxq},  the red line is the result of a linear fit to the data; {\it Right}: BH-DVCS contribution to the unpolarized cross section, $\sigma_{UU}^{\cal I}$, plotted vs $\phi$. The red line was obtained using the CFF data extracted from the linear fit shown on the left. }
    \label{fig:BHDVCS}
\end{figure*}

Noting the following correspondence between the elastic form factor terms in Eq.(\ref{eq:BH}), and the various terms in Eqs.(\ref{eq:siguuI},\ref{eq:sigLUI}),
\begin{eqnarray}
\label{eq:GE}
F_1^2 + \tau F_2^2 & \rightarrow & F_1 \mathcal{H} 
+ \tau F_2 \mathcal{E} \\
\label{eq:GM}
(F_1+F_2)^2 & \rightarrow & G_M (\mathcal{H} + \mathcal{E})  
\\
G_M G_A& \rightarrow  & G_M \widetilde{\mathcal{H}} ,
\end{eqnarray}
one can identify a Rosenbluth formula for the unpolarized DVCS data.
In a Rosenbluth separation for elastic $ep$ scattering one makes cross section measurements at a fixed four momentum transfer squared value (denoted here as $t$) for different values of 
 $\epsilon$ which, in turn,  depends on the electron scattering angle $\theta$ ($\epsilon=1$ for forward scattering, and
$\epsilon$ = 0 for 180$^o$, scattering). Rosenbluth separations of $G_E$ and $G_M$ have enabled precise determinations of the nucleon form factors in the $t < 1$ GeV$^2$ region, from the 1960's to present. Here we show that a similar goal to obtain separately the CFFs ${\cal H}$ and ${\cal E}$, is at reach for the DVCS experimental analyses.

To make the correspondence between the BH cross section, Eq.(\ref{eq:BH}), and elastic $ep$ scattering even clearer, we rewrite it in terms of contributions from the longitudinally and transversely polarized virtual photon as,
\begin{eqnarray}
\label{eq:siguuBH}
\sigma_{UU}^{BH} = \Gamma  \; \frac{A_{BH} - B_{BH} (1+\tau)}{1+\tau}   \Big[\epsilon_{BH} G_E^2 +  \tau G_M^2  \Big]
\end{eqnarray}
with,
\begin{eqnarray}
\label{eq:epsilonBH}
\epsilon_{BH} & = & \left( 1+ \frac{B_{BH}}{A_{BH}}(1+\tau) \right)^{-1} .
\end{eqnarray}
$\epsilon_{BH}$ measures the exchanged virtual photon's longitudinal polarization relative to the transverse one in the BH process. 
\begin{figure}
    \centering
    \includegraphics[width=6cm,height=12cm]{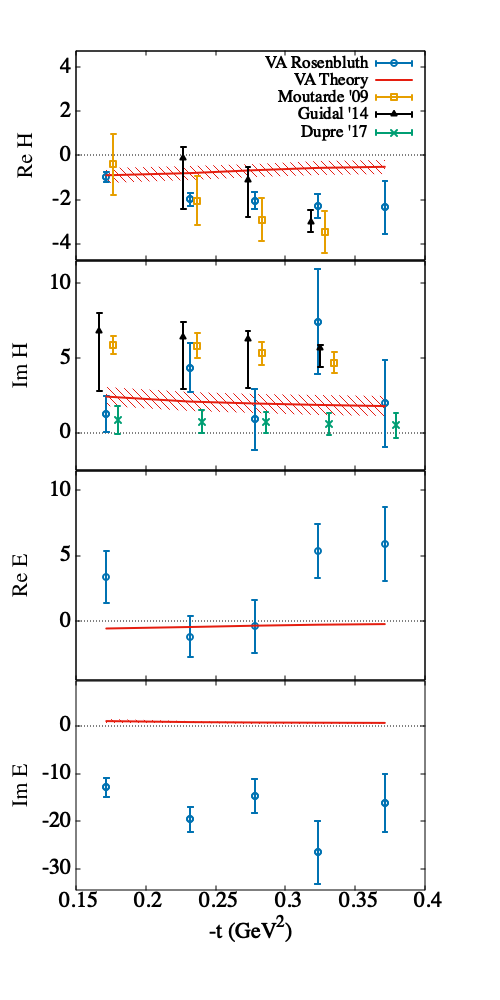}
    \caption{CFFs $\Re e {\cal H}$, $\Im m {\cal H}$, $\Re e {\cal E}$ and $\Im m {\cal E}$ (from top down) extracted from the Rosenbluth separation in this paper (blue circles), and the analyses of Ref.\cite{Guidal:2008ie}  (black triangles) , Ref.\cite{Moutarde:2009fg}  (yellow squares), Ref.\cite{Dupre:2017hfs} (green crosses) . The curves are predictions from the model parametrization of Ref.\cite{GonzalezHernandez:2012jv}.}
    \label{fig:CFFresults}
\end{figure}

In a Rosenbluth separation for the BH process, we fix $t$ and vary the value of 
$\epsilon_{BH}$, which now depends on both $Q^2$ (related to $\theta$) and $\phi$. 
In Figure \ref{fig:BH} we show Rosenbluth separated $ep \rightarrow e' p' \gamma$ data in one of the kinematic bins from Ref.\cite{Defurne:2015kxq}. 
\begin{figure}
    \centering
    \includegraphics[width=7cm]{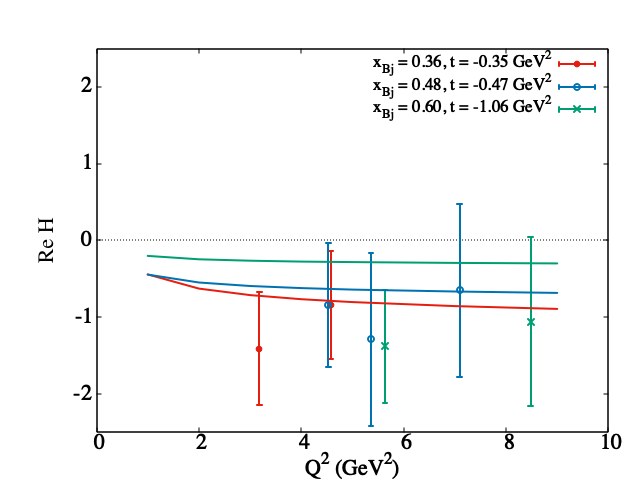}
    \caption{$Q^2$ dependence of the reduced cross section, $\sigma_{UU}^{\cal I} Q^2 | t | /\Gamma$, Eq.(\ref{eq:siguuI}), in the kinematic bins from Ref.\cite{Georges:2018kyi}: $x_{Bj} = 0.36$, $t = -0.35$ GeV$^2$ (red full circles);  $x_{Bj} = 0.45$, $t = -0.47$ GeV$^2$ (blue open circles); $x_{Bj} = 0.60$, $t = -1.06$ GeV$^2$ (green crosses). The curves correspond to a prediction obtained evolving in LO perturbative QCD the CFFs in the model calculation Ref.\cite{GonzalezHernandez:2012jv}. The kernels for LO evolution were first obtained in Refs.\cite{Ji:1996nm, GolecBiernat:1998ja}.}
    \label{fig:Q2}
\end{figure}
On the left panel we plot the reduced cross section, $\sigma_{UU} t^2/\Gamma B_{BH}$ vs. the ratio $A_{BH}/B_{BH}$, related to $\epsilon_{BH}$ in Eq.\eqref{eq:epsilonBH}. The calculation represents the reduced BH cross section, $\sigma_{BH} t^2/\Gamma B_{BH}$, which, according to Eqs.(\ref{eq:BH},\ref{eq:siguuBH}), is a straight line intercepting the $y$-axis at $\tau G_M^2$. On the right panel we show the same data plotted vs. $\phi$, in the standard way. One can clearly see the deviation of the data from the calculation of the BH cross section due to the presence of DVCS.

In order to extend a similar method to the BH-DVCS interference cross section, 
 we first extract $\sigma_{UU}^{\cal I}$ from the total cross section data by subtracting the exactly calculable $\sigma_{BH}$ term, and by evaluating the pure DVCS term from the intercept at $\phi=90^o$, and $\phi=270^o$. 

%
In Figure \ref{fig:BHDVCS} we show results for the same experimental bin as Fig.\ref{fig:BH}: on the {\it l.h.s.} we show the reduced BH-DVCS interference cross section, $\sigma_{UU}^{\cal I} Q^2 t/(\Gamma B^{\cal I})$ plotted vs. $A^{\cal I}/B^{\cal I}$, while on the {\it r.h.s} we show the cross section, $\sigma_{UU}^{\cal I}$, Eq.(\ref{eq:siguuI})  plotted vs. $\phi$. 
Notice the advantage in transitioning from the complex $\phi$ functional dependence on the {\it r.h.s.}, to the linear form on the {\it l.h.s.}. 
%

Demonstrating the possibility of simultaneously extracting all four form factors, in a model independent way, while the contribution of $\widetilde{{\cal H}}$ is suppressed is the main result reported in this Letter. 

We note that the errors in Fig.\ref{fig:BHDVCS} are larger than in the BH case. We attribute this to two main reasons: the data become noisier once the dominant BH term is subtracted, and the Rosenbluth extraction becomes less precise when the interference contribution approaches zero at around $\phi=90^o$ (central two points). In addition, according to the new formalism, the coefficient, $C^{\cal I}$, of the axial vector CFF, $\widetilde{\cal H}$,  is negligibly small compared to the other coefficients. 

The same procedure can be applied to the polarized electron data, Eq.(\ref{eq:sigLUI}), yielding all four CFFs, $\Re e {\cal H}$, $\Im m {\cal H}$, $\Re e {\cal E}$, $\Im m {\cal E}$. 
The latter are shown in Figure \ref{fig:CFFresults}, plotted vs. $t$ at fixed $Q^2= 2$ GeV$^2$, $\varepsilon_1 = 6$ GeV, and $x_{Bj}=0.36$, corresponding to the kinematic bin measured in  Ref.\cite{Defurne:2015kxq}. Our results are consistent with  previous extractions of $\Re e {\cal H}$ from Refs. \cite{Moutarde:2009fg,Guidal:2008ie} , and of $\Im m {\cal H}$ from Ref.\cite{Dupre:2017hfs}. We show for the first time the extracted values of $\Re e {\cal E}$ and $\Im m {\cal E}$ from data.     

Extending our analysis to more recent data obtained at Jefferson Lab @ 12 Gev \cite{Georges:2018kyi} affords us to look at the $Q^2$ dependence of the CFFs. As shown in Figure \ref{fig:Q2}, the values of the CFFs extracted in the new formalism are consistent with the predictions from perturbative $Q^2$ evolution of GPDs (\cite{Ji:1996nm} and \cite{GolecBiernat:1998ja}). Although more data are needed to map out precisely the behavior of the CFFs with $Q^2$, this result can be interpreted as a confirmation that twist three and higher order $Q^2$-dependent effects are small within error, contrarily to what claimed in Ref.\cite{Defurne:2017paw}. 

We checked how the contribution of the twist three terms weighs in, in our analysis.
These were evaluated in \cite{Kriesten:2019jep,Kriesten:2020wcx}. For the unpolarized target case we have the additional contribution to $\sigma_{UU}^{\cal I}$, Eq.(\ref{eq:sigLUI}),
\begin{eqnarray}
\label{eq:Int_FUU3}
&& \frac{\sqrt{t_o-t}}{Q^2}   \Big\{ A^{ (3) \cal I}   \Big[ F_{1} \Big( {\cal H}_\perp - \widetilde{\cal H}_\perp \Big)
 +  F_{2}  \Big( {\cal H}_T^{(3)} - \widetilde{\cal H}_T^{(3)} \Big) \Big] 
 \nonumber \\
 &+ &  B^{(3) \cal I} G_M \,  (  {\cal H}_L^\perp - \widetilde{\cal H}_L^\perp ) \nonumber 
 \\
 &+ &  C^{(3) \cal I}  G_M \,    \Big[2\xi ( \mathcal{H}_{2T} -  \mathcal{H}_{2T}')- \tau \Big( \  {\cal H}_L^\perp - \widetilde{\cal H}_L^\perp  \Big)\Big] \Big\},
\end{eqnarray}
where $t_o$ is the minimum kinematically allowed value of $t$, and the expressions for the longitudinally polarized electron, are analogous to Eq.(\ref{eq:Int_FUU3}), replacing $\Re e \rightarrow \Im m$. The twist three GPDs are defined in Ref.\cite{Kriesten:2020wcx} where a comparison with the notation of Ref.\cite{Meissner:2009ww} is also given. The size of the coefficients $A^{(3) \cal I}$, $B^{(3) \cal I}$ and $C^{(3) \cal I}$, which were evaluated exactly in Ref.\cite{Kriesten:2019jep}, turns out to be small, and comparable in size to the axial vector coefficient, $C^{\cal I}$. These coefficients are, therefore, absorbed in the systematic error of the twist-two terms. We conclude that the extraction of twist-three observables will have to rely on ``super-observables" with specific polarization configuration combinations, or combinations of results from different experiments, namely DVCS and timelike Compton scattering \cite{Boer:2015fwa,Boer:2015cwa}, such that the relative contribution of the twist two terms is suppressed.

To summarize, we emphasize that our quantitative analysis represents a proof of concept opening the way for more detailed and systematic studies to scan the behavior of the Rosenbluth separation variables, {\it e.g.}  $A^{\cal I}/B^{\cal I}$, as a function of the kinematic variables $\varepsilon_1$, $Q^2$, the scattering angle, $\theta$, and the azimuthal angle, $\phi$. Similar to elastic scattering, future experimental measurements could attain a precise determination of all CFFs by refining the kinematic coverage needed to produce the linear plots proposed in our study. 

Given the complicated structure of the DVCS cross section, uncovering linear relations among its various components provides a remarkable simplification of the formalism and a pathway to a straightforward data interpretation. 
For the case illustrated here, linear relations enable a Rosenbluth separation technique to simultaneously extract the values of the CFFs $\mathcal{H}$ and $\mathcal{E}$ from the unpolarized target cross section. 
The benefits of this analysis, however, do not stop at this step:  our formalism, by introducing a clear description of the electric, magnetic and axial components, highlights the possibility of using complementary methods to optimize the extraction of CFFs from experiment. These include double polarization measurements for proton recoil polarization, $\overrightarrow{e} p \rightarrow e' \overrightarrow{p}' \gamma$, and polarized scattering, $\overrightarrow{e} \overrightarrow{p} \rightarrow e' p' \gamma$. Developing separation techniques for DVCS and related processes, along the lines we propose would providesconstitute 
and an essential backdrop for explorations using Machine Learning based algorithms \cite{Cuic:2020iwt,Our_ML}  and future state of the art methodologies in both data analysis and visualization.  
 


\acknowledgements
This work was funded by DOE grants DOE grant DE-SC0016286 and in part by the DOE Topical Collaboration on TMDs (B.K. and S.L.), and SURA grant C2019-FEMT-002-04. We gratefully acknowledge discussions with Matthias Burkardt, Gordon Cates, and Xiangdong Ji.

\bibliography{DVCS_BH_bib}

\end{document}